\documentclass{article}

\usepackage{arxiv}

\usepackage[utf8]{inputenc} 
\usepackage[T1]{fontenc}    
\usepackage{hyperref}       
\usepackage{url}            
\usepackage{booktabs}       
\usepackage{amsfonts}       
\usepackage{nicefrac}       
\usepackage{microtype}      
\usepackage{lipsum}
\usepackage{graphicx}
\usepackage[most]{tcolorbox} 
\graphicspath{ {./images/} }
\usepackage{url} 
\usepackage{tabularx}
\usepackage{balance}
\usepackage{adjustbox}
\usepackage{color}
\usepackage{fancybox}
\usepackage{graphicx}
\usepackage{cellspace}
\usepackage{comment}
\usepackage{diagbox}
\usepackage{url}
\usepackage{xspace}
\usepackage{amssymb}
\usepackage{multirow}
\usepackage[caption=false]{subfig}
\usepackage{epsfig}
\usepackage{cellspace}
\usepackage[T1]{fontenc}
\usepackage{amsmath}

\newtcbtheorem{Summary}{\bfseries Hypothesis}{enhanced,drop shadow={black!50!white},
  coltitle=black,
  top=0.2in,
  attach boxed title to top left=
  {xshift=1.5em,yshift=-\tcboxedtitleheight/2},
  boxed title style={size=small,colback=gray!40}
}{summary}

\newtcolorbox{box2}[1][]
{
  title=#1, 
  enhanced,
  colback=gray!20,
  frame hidden,
  colbacktitle=gray, 
  boxed title style={colframe=gray},
  top=0.05cm,
  bottom=0.05cm,
  left=0.15cm,
  right=0.15cm,
  enlarge top by=0.5cm,
  enlarge bottom by=0.3cm,
  attach boxed title to top left={xshift=3mm,
  yshift=-3mm,yshifttext=-1mm},
  borderline west={4pt}{0pt}{blue!50!black},
  breakable
}
\title{Architecting Peer-to-Peer Serverless Distributed Machine Learning Training for Improved Fault Tolerance}

\author{ 
{Amine Barrak} \\
	Department of Computer Science\\
	Université du Québec à Chicoutimi\\
	Saguenay, QC \\
	\texttt{mabarrak@etu.uqac.ca} \\
	\And
	{Fabio Petrillo} \\
	Department of Software Engineering\\
	École de Technologie Supérieure\\
	Montreal, QC \\
	\texttt{fabio@petrillo.com} \\
	\And
	{Fehmi Jaafar} \\	Department of Computer Science\\
	Université du Québec à Chicoutimi\\
	Saguenay, QC \\
	\texttt{fehmi.jaafar@uqac.ca} 
}
\date{} 	
\begin{document}
\maketitle
\begin{abstract}
Distributed Machine Learning refers to the practice of training a model on multiple computers or devices that can be called nodes. Additionally, serverless computing is a new paradigm for cloud computing that uses functions as a computational unit. Serverless computing can be effective for distributed learning systems by enabling automated resource scaling, less manual intervention, and cost reduction. By distributing the workload, distributed machine learning can speed up the training process and allow more complex models to be trained. Several topologies of distributed machine learning have been established (centralized, parameter server, peer-to-peer). However, the parameter server architecture may have limitations in terms of fault tolerance, including a single point of failure and complex recovery processes.
Moreover, training machine learning in a peer-to-peer (P2P) architecture can offer benefits in terms of fault tolerance by eliminating the single point of failure. In a P2P architecture, each node or worker can act as both a server and a client, which allows for more decentralized decision making and eliminates the need for a central coordinator. In this position paper, we propose exploring the use of serverless computing in distributed machine learning training and comparing the performance of P2P architecture with the parameter server architecture, focusing on cost reduction and fault tolerance.
\end{abstract}

\section{Introduction}
\label{sec:introduction}
Machine learning is a rapidly growing field that requires high computing resources, particularly when training large, complex models. As the demand for precise and advanced machine learning models rises, the importance of having high computing resources has become increasingly crucial. Moreover, training ML models is a data-intensive activity that faces large-scale computing issues with sophisticated models and data lakes, and data input may become a severe performance bottleneck \cite{haussmann2018accelerating}. Given the substantial computing requirements of machine learning, distributed training has emerged as a necessary approach that enables multiple nodes to share their computing resources.

Cloud computing has revolutionized the way machine learning models are trained and deployed. With cloud computing, high-performance computing resources, such as GPUs and TPUs, are available, which can significantly speed up the training process. Additionally, cloud-based platforms offer parallelization capabilities, enabling powerful model training across multiple machines \cite{hwang2017cloud}. In particular, serverless is a new paradigm for cloud computing that uses functions as the unit of computation. It simply requires the function code to be uploaded, and the cloud provider will take care of running these functions, including managing resources such as servers and storage. These functions can be triggered by specific events, such as an HTTP request, a message in a queue, or a change in a database. They are designed to perform a specific task and are executed only when needed, making them highly scalable and cost-effective. Serverless functions are typically stateless, and once the function has completed its task, it is terminated, freeing up resources for other functions to use \cite{schleier2021serverless}.

Distributed Machine Learning refers to the practice of training a machine learning model on multiple computers or devices that can be called nodes. This is done to scale up the training process, handle large amounts of data, and leverage the combined computing power of multiple machines. By distributing the workload, distributed machine learning can speed up the training process and allow for larger and more complex models to be trained. Several topologies of distributed machine learning has been set (centralised, parameter server, peer to peer), where The different nodes of the distributed system need to be connected through a specific architectural pattern to fulfill a common task \cite{verbraeken2020survey}.


The degree of distribution that the system is planned to implement is a decisive element in topology. It can have a substantial impact on its performance, scalability, dependability, and security \cite{verbraeken2020survey}. Specifically, distributed learning is a method of training machine learning based on a leader-worker architecture, where multiple worker machines work together under the supervision of a leader machine to train a model. Several solutions for distributed machine learning based on parameter server architecture using serverless were proposed \cite{P38, P39, P46, P51, P53, P54}. In a parameter server architecture, each worker updates its own model parameters locally and periodically sends its updates to the parameter server. The parameter server aggregates these updates and broadcasts them to all workers. This allows for efficient parallel training of the model, as each worker can continue training using the latest version of the model parameters.


However, the parameter server architecture can have limitations in terms of fault tolerance, including a single point of failure, server communication overhead, complex recovery processes, and lack of redundancy of the parameter server, as well as load balancing \cite{addairdecentralized}.

Byzantine faults refer to any arbitrary or unpredictable behavior of a component in the system. In distributed machine learning training, a Byzantine node can affect the training process of a model and lead to incorrect or compromised results. Guerraoui et al. \cite{guerraoui2021garfield} conducted an experiment to test the additional overhead of fault tolerance in different architectures based on servers and workers. They found that tolerating Byzantine servers induces much more overhead than tolerating Byzantine workers.

Fault tolerance is essential in machine learning, especially in distributed systems with multiple nodes or devices involved in model training. A failure in one component can have significant impacts on the system's overall performance, leading to wasted time, resources, and money. By implementing fault tolerance measures, machine learning systems can improve their reliability and minimize the risk of costly downtime and data loss, resulting in more efficient use of resources and better outcomes \cite{myllyaho2022misbehaviour}.

Peer-to-peer architectures are more resilient and fault-tolerant than parameter-server architectures because there is no central server that is prone to failure (for example, during high demand), avoiding the well-known problem of a Single Point of Failure (SPOF) that can bring an entire machine learning system to a halt. In a P2P architecture, each node or worker can act as both a server and a client, which allows for more decentralized decision-making and eliminates the need for a central coordinator.

Serverless computing offers several computing capabilities, such as cost optimization \cite{elgamal2018costless} and high scalability \cite{yu2020characterizing}, that are well-suited to the expensive computing operations of machine learning.


This \textbf{positional paper} presents our hypothesis regarding a peer to peer distributed training based on serverless computing. This hypothesis can provide a valuable insight in term of fault tolerance.

\section{Proposal}
\label{sec:proposal}
In this positional paper, we propose using a peer-to-peer serverless architecture to improve the fault tolerance of distributed machine learning training. Our research addresses this hypothesis in detail, which is discussed next.

\begin{Summary}{}{}
Distributed machine learning training using serverless computing with peer-to-peer (P2P) architecture results in improved fault tolerance compared to using a serverless-based parameter server architecture.
\end{Summary}
\vspace{0.4cm}

To validate our hypothesis, our research project is divided into the following three steps: (1) Implement a peer-to-peer architecture based on serverless computing for distributed machine learning training; (2) Evaluate and explore the impact of serverless computing on the peer-to-peer architecture; (3) Evaluate and compare the fault tolerance of the serverless-based peer-to-peer architecture with that of the serverless-based parameter server architecture.

\section{Conclusion}
\label{sec:conclusion}
In this paper, we present our proposal of exploring the impact of serverless computing on distributed training for machine learning using peer-to-peer architecture. Our goal is to investigate the potential benefits and challenges associated with this approach, with a focus on fault tolerance. To achieve this, our research design for comparing the performance of a serverless-based parameter server architecture and a P2P architecture in distributed training machine learning.

\bibliographystyle{unsrt}  
\bibliography{references}  

\begin{thebibliography}{10}

\bibitem{haussmann2018accelerating}
Elmar Haussmann.
\newblock Accelerating i/o bound deep learning on shared storage, 2018.

\bibitem{hwang2017cloud}
Kai Hwang.
\newblock {\em Cloud computing for machine learning and cognitive
  applications}.
\newblock Mit Press, 2017.

\bibitem{schleier2021serverless}
Johann Schleier-Smith, Vikram Sreekanti, Anurag Khandelwal, Joao Carreira,
  Neeraja~J Yadwadkar, Raluca~Ada Popa, Joseph~E Gonzalez, Ion Stoica, and
  David~A Patterson.
\newblock What serverless computing is and should become: The next phase of
  cloud computing.
\newblock {\em Communications of the ACM}, 64(5):76--84, 2021.

\bibitem{verbraeken2020survey}
Joost Verbraeken, Matthijs Wolting, Jonathan Katzy, Jeroen Kloppenburg, Tim
  Verbelen, and Jan~S Rellermeyer.
\newblock A survey on distributed machine learning.
\newblock {\em Acm computing surveys (csur)}, 53(2):1--33, 2020.

\bibitem{P38}
Marc S{\'a}nchez-Artigas and Pablo~Gimeno Sarroca.
\newblock Experience paper: Towards enhancing cost efficiency in serverless
  machine learning training.
\newblock In {\em Proceedings of the 22nd International Middleware Conference},
  pages 210--222, 2021.

\bibitem{P39}
Andreas Grafberger, Mohak Chadha, Anshul Jindal, Jianfeng Gu, and Michael
  Gerndt.
\newblock Fedless: Secure and scalable federated learning using serverless
  computing.
\newblock {\em arXiv preprint arXiv:2111.03396}, 2021.

\bibitem{P46}
Jiawei Jiang, Shaoduo Gan, Yue Liu, Fanlin Wang, Gustavo Alonso, Ana Klimovic,
  Ankit Singla, Wentao Wu, and Ce~Zhang.
\newblock Towards demystifying serverless machine learning training.
\newblock In {\em Proceedings of the 2021 International Conference on
  Management of Data}, pages 857--871, 2021.

\bibitem{P51}
Daniel Barcelona-Pons, Pierre Sutra, Marc S{\'a}nchez-Artigas, Gerard
  Par{\'\i}s, and Pedro Garc{\'\i}a-L{\'o}pez.
\newblock Stateful serverless computing with crucial.
\newblock {\em ACM Transactions on Software Engineering and Methodology
  (TOSEM)}, 31(3):1--38, 2022.

\bibitem{P53}
Pablo~Gimeno Sarroca and Marc S{\'a}nchez-Artigas.
\newblock Mlless: Achieving cost efficiency in serverless machine learning
  training.
\newblock {\em arXiv preprint arXiv:2206.05786}, 2022.

\bibitem{P54}
Ahsan Ali, Syed Zawad, Paarijaat Aditya, Istemi~Ekin Akkus, Ruichuan Chen, and
  Feng Yan.
\newblock Smlt: A serverless framework for scalable and adaptive machine
  learning design and training.
\newblock {\em arXiv preprint arXiv:2205.01853}, 2022.

\bibitem{addairdecentralized}
Travis Addair.
\newblock Decentralized and distributed machine learning model training with
  actors.

\bibitem{guerraoui2021garfield}
Rachid Guerraoui, Arsany Guirguis, J{\'e}r{\'e}my Plassmann, Anton Ragot, and
  S{\'e}bastien Rouault.
\newblock Garfield: System support for byzantine machine learning (regular
  paper).
\newblock In {\em 2021 51st Annual IEEE/IFIP International Conference on
  Dependable Systems and Networks (DSN)}, pages 39--51. IEEE, 2021.

\bibitem{myllyaho2022misbehaviour}
Lalli Myllyaho, Mikko Raatikainen, Tomi M{\"a}nnist{\"o}, Jukka~K Nurminen, and
  Tommi Mikkonen.
\newblock On misbehaviour and fault tolerance in machine learning systems.
\newblock {\em Journal of Systems and Software}, 183:111096, 2022.

\bibitem{elgamal2018costless}
Tarek Elgamal.
\newblock Costless: Optimizing cost of serverless computing through function
  fusion and placement.
\newblock In {\em 2018 IEEE/ACM Symposium on Edge Computing (SEC)}, pages
  300--312. IEEE, 2018.

\bibitem{yu2020characterizing}
Tianyi Yu, Qingyuan Liu, Dong Du, Yubin Xia, Binyu Zang, Ziqian Lu, Pingchao
  Yang, Chenggang Qin, and Haibo Chen.
\newblock Characterizing serverless platforms with serverlessbench.
\newblock In {\em Proceedings of the 11th ACM Symposium on Cloud Computing},
  pages 30--44, 2020.

\end{thebibliography}


\end{document}